# Query Centric CPS (QCPS) Approach for Multiple Heterogeneous Systems

[1]**Ankit Mundra**, [2]**Nitin Rakesh**, [3]**Vipin Tyagi**
[1,2]Dept. of CSE, Jaypee University of IT, Waknaghat, Solan, HP, India
[3]Dept. of CSE, Jaypee University of Engg. and Technology, Raghogarh, MP, India

## Abstract
In modern scenario we need to have mechanisms which can provide better interaction with physical world by an efficient and more effective communication and computation approach for multiple heterogeneous sensor networks. Previous work provides efficient communication approach between sensor nodes and a query centric approach for multiple collaborative heterogeneous sensor networks. Even there is energy issues involved in wireless sensor network operation. In this paper we have proposed Query-centric Cyber Physical System (QCPS)model to implement query centric user request using Cyber Physical System (CPS). CPS takes both communication and computation in parallel to provide better interaction with physical world. This feature of CPS reduces system cost and makes it more energy efficient. This paper provides an efficient query processing approachfor multiple heterogeneous sensor networks using cyber physical system.This approach results in reduction of communication and computation cost as sensor network communicates using centroid of respective grids which reduces cost of communication whileinvolvement of CPS reduces the computation cost.

## I. Introduction and Motivation
Sensor network is on demand research area which fulfills industry need for smart decisive application. To maintain effective and efficient industrial outcome it is required to develop sensor based approaches more effectively in terms of cost and energy. Although it is easy to handle homogeneous sensor networks as they contain identical nodes in terms of battery energy and hardware architecture while in multiple heterogeneous sensor networks all sensor nodes are different in terms of these properties: battery energy and functionality. In the scenario where user need a single result from whole system than it is required to manage the collaborative nature of sensors i.e. dependability on each other as the communication between sensors may cause high energy drainage.

Further it is proposed in previous work Energy efficiency which is main objective ofwireless sensor networks as the nodes are severely energy constrained, and battery replenishment is not practical [1]. To resolve these issues it is required to have some algorithms which can provide minimum energy drainage approach of communication. Our proposed model: QCPS is a model based on cyber physical system and it provides communication approach for the sensors by arranging in the form of grid. This model enables faster query execution with low energy consumption in collaborative heterogeneous sensor networks.

### A. Introduction of Cyber Physical System
Cyber physical system is the new generation architecture which incorporates computation and communication with physical control process.Now a day'sCPS is viewed as a new science for future engineered. It is a system with controlled physical integrated communication and computational capabilities [2]. Because of advancementin modern technology where physical interaction needscyber physical system gets the attention of researcher and can be viewed for both developments in academic and industrial field.

CPS research is still in its early stage i.e. research is partitioned into many categories and sub disciplines such as sensors, communication and networking, control theory, software engineering and computer science. CPSis expected to play a major role in the design and development of future engineering systems with the new capabilities for exceed today's level of autonomy, functionality, usability, reliability and cyber security.Advances in CPS research can be accelerated by close collaborations between academic discipline in computation, communication, control and other engineering and computer science discipline, coupled with grand challenge applications [3]. The major Applications of CPS include high confidence medical devices and systems, traffic control and safety which detect traffic accidents and provide situational awareness [4], advanced automotive systems, monitor cardiac patients [5-6], congestion controlin real time systems [7].

fig. 1 shows the components of CPS as: It has tightly coupled communication and computation with the control part of system. CPS systems facilitate better functionality as they provide intensive interaction of physical and computational componentsand due to these computational components are aware of physical context of system and they are well synchronized.

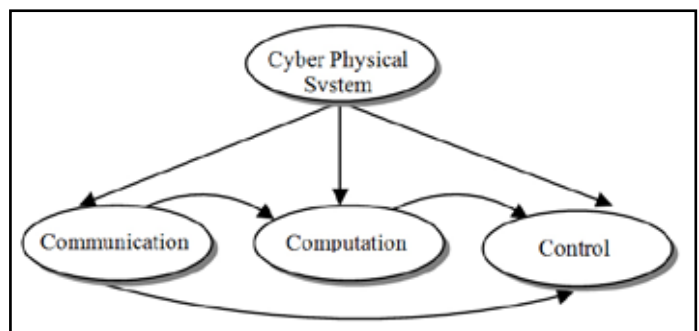

Fig. 1: Cyber Physical Systems

This paper is divided in five sections. In sectionfirst the problem is defined and the concept related to the problem is introduced withintroduction of cyber physical system. In section two we briefly definedproposed architecture for our problem. Further in section third we illustrate the working of our proposed model i.e. working of QCPS model. In section four we see the based on our model. Finally we conclude in section five.

## II. QCPS Architecture
To introduce our approach: QCPS, we have proposed its architecture so that its work can be classified for the user prospective. The QCPS architecture is proposed for multiple heterogeneous sensor networks. This architecture overcomes the issues of numerous sensors for communication and computation; collaborative nature of sensor nodes, and provides a generic mechanism for sensor communication in distinct databases in cloud for unique user request using CPS.

Fig. 2 shows the QCPS architecture. It consist three inter-dependable components: first one comprises collaborative heterogeneous networks i.e. the network are depend on one another. Networks sensed the data and send it to cloud for storage and computation. Second one, which is also known as storage and computation unit is the cloud. In cloud, it maintains separate database for each type of network.

Cloud receives data from networks and stores them into their respective database, where it also performs computation on stored data. In cloud there is a network recognize engine which identifies the networks and classified their data according the network type and send to respective database. In the QCPS model as we use clouds because of following advantages [8-9]:

- Rapid Elasticity: Clouds provide rapid elasticity so it can scale the resources as per user's need.
- On demand self-service: It provides on demand service that is why resource use is optimized.
- Everywhere network access: Cloud provide the access to the user regardless from location of user.
- Pricing: It is not very expensive to store data in the cloud and then again access from it, so it does not have any upfront cost the cost is totally based on usage.
- Quality of service: clouds maintain Quality of service.
- Security: only authorized users in one account can access data from it.

Third one is the group of user. This architecture enables the users to directly interact with cloud for the centric query processing.

QCPS architecture makes use of cyber physical system as: It performs concurrent communication and computation i.e. at the same time sensors sense their medium and present the data for computation and after computation data is available for sharing. This architecture provide the mechanism for interaction with two issues i.e. first membership management done by formalizing the sensor node which is discussed in next section. Second Information sharing between heterogeneous network is done by cloud i.e. each time when sensors need data of other sensors then it can extracted from cloud. Further the working of QCPS model will be briefly discuss in next section.

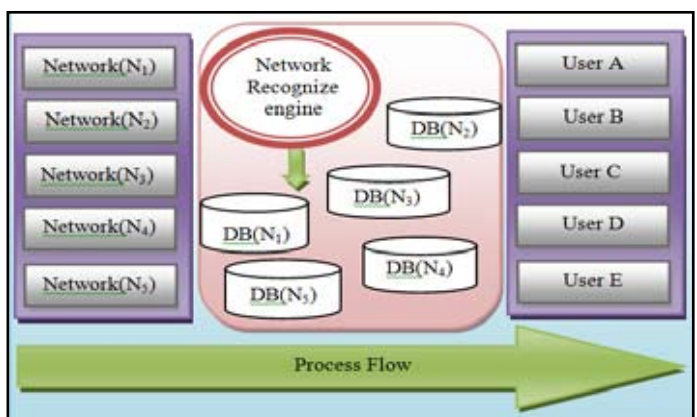

Fig. 2: QCPS Architecture

### III. QCPS Model

This section describes the working of QCPS model in detail. For the query centric approch in Multiple Heterogeneous sensor networks, we have interdependable networks so that we need a model which can provide a efficient way to support the heterogeneous enviornment.

Before start the working of QCPS model we need some assumptions as: here we broadly classify the sensors by their primary input quantity (measured), transduction principles, material and technology, application [10]. They can classify into their sub types for e.g. Environment sensors is in of temperature, humidity, light sensors etc. and activity sensors are in human activity, object activity etc. and electric sensors are in charge, current, voltage, conductivity sensors etc. and mechanical sensors are in position, acceleration, force, stress, pressure, torque, shape, orientation, stiffness, compliance sensors etc.

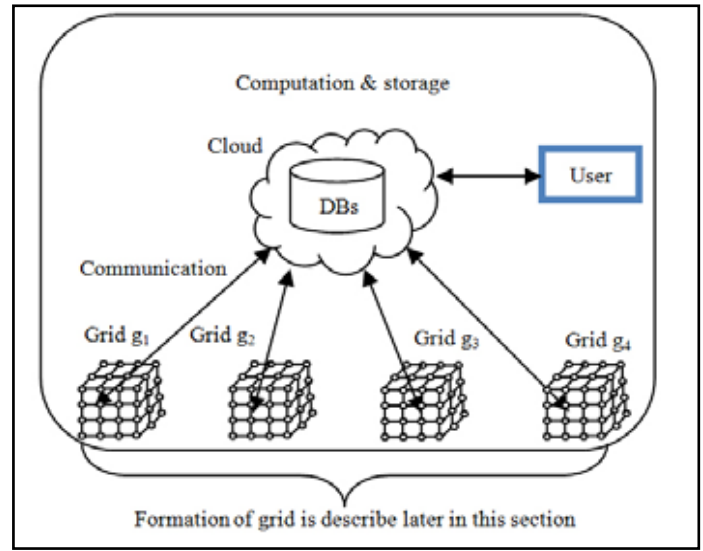

Fig. 3: Basic Scenario of QCPS Model

Fig. 3 shows a basic scenario of QCPS model as: It contains the four different sensor networks each arranged in the form of grid and cloud where DBs are maintain for each network. In this communication is performed between grids and cloud. On the other hand computation is performed at cloud only. By taking the above scenario we define the general QCPS model as:

1. Same kind of sensors arranged in grid form by using 3D mesh structure because mesh topology has its advantages as:
- It forms regularly distributed network
- Its provide peer to peer connectivity between nodes
- Although all nodes may be identical and have the same computing and transmission capabilities, certain node can be designated as 'Coordinator node' that take on additional functions. If a coordinator node is disabled, another node can then take over these duties

2. For each grid there is a coordinator node. This makes communication with cloud for storing the data of the members of its own grid as well as extracting the data from databases. So when a sensor needs data of other sensors then it makes a request to coordinator node.

3. In Cloud there is a separate table for each network(e.g. activity sensors have a different database and environment sensors have their separate database).

4. Whenever user has a central query then he/she can directly interact with cloud and get the response.

In this approach the benefits are: Firstly computation is performed in cloud so that the energy drainage in sensors reduces, because they don't need to perform computation any more. Secondly communication cost also reduces because coordinator can access the data from cloud and deliver it to demanding nodes so that sensor node does not have to communicate with the node situated very far.
So we can understand the working of model by taken an example: Sensor (for e.g.A) of particular network ($N_1$) needs the data of another sensor (B) which resides in another network ($N_2$), than the Sensor (A) go to its coordinator node (C1) and then that coordinator node ($C_1$) get the data of (B) from cloud anddeliver it to Sensor (A).

### A. Formalization of Sensor Node

In Query Centric Approach for multiple collaborative heterogeneous systems where each sensor node want to communicate with another sensor node [11], it becomes important to formalize each node with some attributes. Also for membership management of each sensor, we need to formalize the sensor node as:

$$Node_i = [Node\ id, Type, Grid\ id, Physical\ Location, Coordinator(Boolean)]$$

These attributes are defined as in Table 1:

Table 1: Description of QCPS Sensor Attributes

| Attributes | Description |
|---|---|
| Node id | Uniquely defines each sensor in the system. |
| Type | Type of sensors (E.g.: activity, environment, biological etc.) |
| Grid id | Uniquely defines each grin in the system. |
| Physical Location | Location where sensor is deployed. |
| Coordinator | True for Coordinator node and false for remaining nodes. |

For authentication and identification of the node which is trying to access data from databases, network recognize engine fig. 3verifies the attributes value of the node.And after validates the node it is allows to extract data from databases.
Algorithm to Compute Grid
This algorithm computes a grid from the sensor networks. It takes all heterogeneous sensors as input and assigns them into a grid by checking two conditions as: (1) they should be of same type (2) distance between two sensors must be less than the threshold distance ('Distance').

| |
|---|
| Initial Condition: Different types of sensors |
| Final Condition: |
| Given: Types of Sensors and Physical Location |
| Local variables: Distance, Type, i, j,N |
| Initialize Local Variables i,j,N with zero |
| *For* all node N=1,2,…,n |
| *For* i=1 to n |
| *For* j=i+1 to n |
| If |
| /* It checks two conditions first is type of two sensor nodes and second is distance between two nodes should less than Pre define Distance*/ |
| **Then** |
| 　　　They are assigning to a same Grid |
| **Else** |
| Repeat |
| |
| Return |

*Note*: *Distance is pre-decided or it has some threshold value in which only sensor can assign to a same grid.*

Type of sensor networks (e.g. Activity, Physical, and Biological etc.)

### C. Approch to Decide Coordinator Node

To elect the coordinator node for a grid we need positions of each node as shown in fig. 4. Then by using equation (1), (2), (3)of geometry we can compute the centroid coordinateand then the node which has least distance from that centroid point will be the coordinator node of that particulargrid.

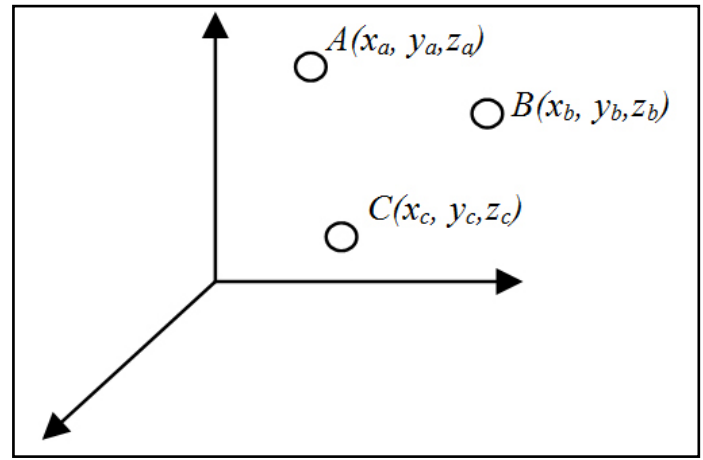

Fig. 4: Arrangement of Sample Nodes in Space

Then we perform following steps to determine the coordinator node within a grid:
• Defineposition of each node as input.
• Compute the centroid by using followingequation:

$$X_c = \sum_{i=1}^{k} \frac{x_i}{k} \quad (1)$$

$$y_c = \sum_{i=1}^{k} \frac{y_i}{k} \quad (2)$$

$$z_c = \sum_{i=1}^{k} \frac{x_i}{k} \quad (3)$$

3. Then compute the nearest node from the centroid by using Euclideandistance formula.
4. Select this node as the coordinator node of respective grid.
5. Return.  /* coordinator node */

### D. How to Store and Access Data from Cloud

In collaborative environment where each sensor node demands data from distinct networks (grids), then QCPS facilitates data requirement through the coordinator of respective grids to communicate via cloud. Fig. 5 shows data sharing between two grids ($g_1$ and $g_2$) which is done by coordinators ($C_1$ and $C_2$) respectively. Further this section describes the steps involved in data sharing process between cloud and grids of QCPS as in Table 2.

Table 1: Steps Involved in Communication and Computation Process

| Steps | Description |
|---|---|
| Step 1 | The required data is shared within two grids (as in example $g_1$ and $g_2$) firstly to respective coordinator node ($C_1$ and $C_2$). |
| Step 2 | Further coordinator nodes $C_1$ and $C_2$ will store their data in respective cloud databases $DB_1$ and $DB_2$ (where computation is performed). |
| Step 3 | Request of individual grids are responded by respective coordinator nodes. |
| Step 4 | Now these coordinator nodes will access respective cloud database to handle the request. |
| */ Each grid elects a coordinator node which will be responsible to handle all request from any node within respective grids and cloud database  /* | |

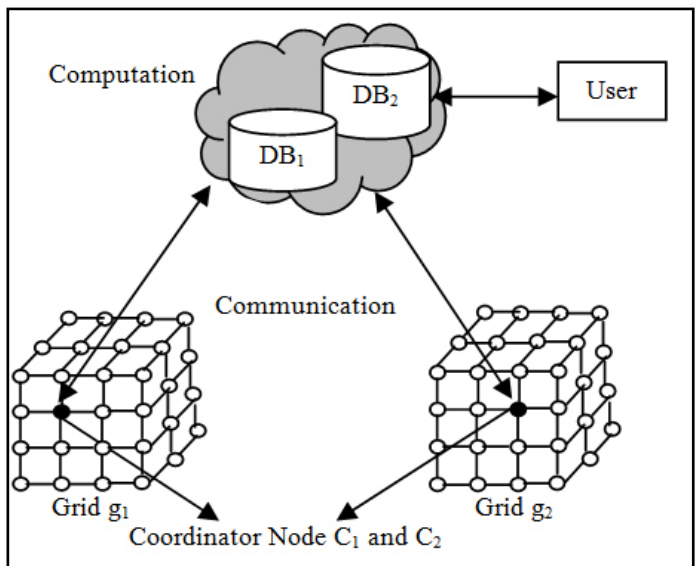

Fig. 5: Communication and Computation Process

## IV. Result

This section provides result of our model. We take a scenario to study and evaluate performance of QCPS model. For this we have considered eight sensors with their respective types and physical locations. The following data shows the distinct types of sensors. We have considered (A, B…H) for different coordinates representing their location:

Here,
$A = \{Environment, (30,40,10)\}$
$B = \{Activity, (40,20,70)\}$
$C = \{Activity, (50,70,120)\}$
$D = \{Environment, (70,80,100)\}$
$E = \{Environment, (80,15,110)\}$
$F = \{Activity, (65,60,150)\}$
$G = \{Activity, (130,70,160)\}$
$H = \{Environment, (72,78,90)\}$

Now, according to QCPS model firstly we compute grid which comprises of similar kind of sensors. The subsequent section shows the performance of the proposed algorithms when they are subjected to above location and types.

### 1. According to Algorithm (Section III-B)
Here we have used (Distance is decided by the deplorer).

| |
|---|
| For |
| */ When N=1 so take sensor A, it is assign to $g_1$[Type]=A[Type]) /* |
| For |
| For |
| Sensor *B* compared with *A* and because they are different type so *B* is assign to $g_2$. |
| /*a new grid $g_2$[Type] =B[type] created */ |
| Go to next sensor and repeat this process and after repeating it at |
| and distance |
| /* $N_4$-$N_1$ is Distance between them */ |
| So Sensor *C* is assign to $g_1$ |
| At  for sensor *E* there is {$N_1$[]=$N_5$[]} but $_{51}$ so *E* is assign to different grid $g_3$. /* New Grid $g_3$[type]=E[type] */ |
| Repeat this process until all the sensors assign to a grid. |
| Return $_{1234}$ */ |

2. Now this step computes the coordinator nodes for each grid. For1)/*by using the Algorithm (Section III- C) */.
Where A=(30,40,10)  D=(70,80,100)  H=(72,78,90)
So the centroid of Grid $g_1$→Xc=57,Yc=66,Zc=67 so according to nearest neighbor Search the node H will be the nearest node to centroid so it will be the coordinator node for grid $g_1$. Similarly, for grid $g_2$, C will be the coordinator node and for $g_3$ and $g_4$, node E and G respectively.
4. Further each node in the grid sends it data to their coordinator node and then coordinator node stores the whole data of its grid into the database.
5. In the scenario where Node A want data of Node B then it perform following task:
- Go to its Coordinator node H.
- Then H communicate with cloud and there network recognize engine verifies H and allows to get data of B form B's database.
- Deliver data to A.

The overall outcome of the QCPS model on above scenario is shown in Table 3. This table shows the cost of communication and computation and their respective effects (result) and the corresponding reason. Thus energy consumption is directly proportional to cost of communication and computation. Using our model these issues are resolved.

Table 3: QCPSModel's Outcome

| Costs | Effect | Reason |
|---|---|---|
| Communication | Reduces | Because a node has to communicate with coordinator node only. |
| Computation | Reduces | Computation is performed at cloud not at the sensor node. |

## V. Conclusion and Future work

In this paper we have implemented Cyber Physical System on multiple heterogeneous sensor networks.This paper shows that by using CPS the issues of energy consumption due to high communication and computation cost get resolved by using proposed QCPS model.To show our result we have considered scenario of eight sensorsand have implemented QCPS model withthese sensors.This resultsin reduced cost of communication and computation which enables reduced energy consumption and faster query processing. As a part of future work all the parameters of QCPS model are required to be implemented physically and the results are evaluated in terms of faster query processing. This model requires extendibilityfor large network applications.

## References

[1] Vivek Mhatre, Catherine Rosenberg,"Homogeneous vs Heterogeneous Clustered Sensor Networks: A Comparative Study", School of Electrical and Computer Eng., Purdue University, West Lafayette, In 47907-1285.
[2] Jin Wang, Hassan Abid, Sungyoung Lee, Lei Shu, Feng Xia, "A Secured Health Care Application Architecture for Cyber-Physical Systems", Vol. 13, No. 3, pp. 101-108, 2011
[3] Edward A. Lee,"Cyber-Physical Systems - Are Computing Foundations Adequate?", Position Paper for NSF Workshop on Cyber-Physical Systems: Research Motivation, Techniques and Roadmap October 16 - 17, 2006.
[4] W. Jones,"Forecasting traffic flow", IEEE Spectrum, Vol. 38, No. 1, pp. 90–91, 2001.
[5] P. Leijdekkers, V. Gay,"Personal heart monitoring and rehabilitation system using smart phones", In Proceedings of the International Conference on Mobile Business. Citeseer, pp. 29, 2006
[6] T. Gao et al.,"Participatory User Centered Design Techniques for a Large Scale Ad-Hoc Health Information System", Proc. First ACM SIGMOBILE Int'l Workshop Systems and Networking Support for Healthcare and Assisted Living Environments (HealthNet), 2007.
[7] A. Benveniste, G. Berry,"The synchronous approach to reactive and real-time systems", Proceedings of the IEEE, 79(9), pp. 1270–1282, September 1991.
[8] David Perera,"The real obstacle to federal cloud computing", Fierce Government IT (2012-07-12).
[9] Mao, Ming; M. Humphrey,"A Performance Study on the VM Startup Time in the Cloud", Proceedings of IEEE 5th International Conference on Cloud Computing, pp. 423-430, 2012.
[10] Middelhoek.S., Noorlog.D.J.W., Sensors and Actuators 2 (1981/82), pp. 29-41.
[11] Yuan He,"COSE: A Query-Centric Framework of Collaborative Heterogeneous Sensor Networks", IEEE transactions on parallel and distributed systems, Vol. 23, No. 9, pp. 1681-1693, September 2012.